\documentclass[11pt]{article}
\usepackage{amsmath,amsfonts,amssymb,bbm}
\usepackage{tensor}
\usepackage[T1]{fontenc}
\usepackage[utf8]{inputenc}
\usepackage{lmodern}
\usepackage{authblk}
\usepackage[disable]{todonotes}
\usepackage{color}
\usepackage{multicol}
\usepackage{cite,varwidth}
\usepackage{empheq}
\usepackage{collref}
\usepackage[margin=3cm]{geometry}
\usepackage{hyperref}
\newcommand*\xbar[1]{%
  \hbox{%
    \vbox{%
      \hrule height 0.5pt 
      \kern0.3ex
      \hbox{%
        \kern-0.0em
        \ensuremath{#1}%
        \kern-0.0em
      }%
    }%
  }%
}

\newcommand\Preind[3]{\vphantom{#3}#1#2#3}
\newcommand\email[1]{\thanks{\href{mailto:#1}{\nolinkurl{#1}}}}
\newcommand\td{\text{d}}

\newcommand\cO[1]{{\cal O}(#1)}

\newcommand{\skyp}{{\cal I}^+}
\newcommand{\skym}{{\cal I}^-}
\newcommand{\bz}{\bar{z}}
\newcommand{\bw}{\bar{w}}

\def\bz{\bar{z}}

\def\bY{\bar{Y}}

\def\pzb{\p_{\bz}}
\def\pz{\p_z}
\def\cY{\mathcal{Y}}

\allowdisplaybreaks[1]

\newcommand{\p}{\partial}
\newcommand{\pb}{\p_{\bz}}
\newcommand{\be}{\begin{equation}}
\newcommand{\ee}{\end{equation}}
\newcommand{\bea}{\begin{eqnarray}}
\newcommand{\eea}{\end{eqnarray}}
\newcommand{\beal}{\begin{align}}
\newcommand{\eeal}{\end{align}}
\def\nn{\nonumber}
\newcommand{\half}{\frac{1}{2}}
\def \ga {\gamma_{z\bz}}
\def \gai {\gamma_{z\bz}^{-1}}

\def \sga {\sqrt{\gamma_{z\bz}}}
\def \bw {\bar{w}}
\def \gawi {\gamma_{w\bw}^{-1}}
\def \gaw {\gamma_{w\bw}}
\def \gawik {\gamma_k^{-1}}
\def \gawk {\gamma_k}
\def \cd {\nabla}
\def \goto {$\Longrightarrow$}
\def \tT {\tilde{T}}

\author[a]{Eduardo Conde\email{econdepe@snu.ac.kr}\,}
\author[b]{Pujian Mao\email{maopj@ihep.ac.cn}\,}
\newsavebox\affbox

\affil[a]{\protect\begin{varwidth}[t]{\linewidth}\protect\centering
School of Physics \& Astronomy and Center for Theoretical Physics,\par
Seoul National University, Seoul 08826, South Korea\protect\end{varwidth}\vspace{1.5ex}}
\affil[b]{\protect\begin{varwidth}[t]{\linewidth}\protect\centering
Institute of High Energy Physics and Theoretical Physics Center for Science Facilities,\par
Chinese Academy of Sciences, 19B Yuquan Road, Beijing 100049, P. R. China\protect\end{varwidth}\vspace{1.5ex}}

\title{\bf BMS supertranslations and\\not so soft gravitons}
\date{}

\begin{document}
 \maketitle
 \thispagestyle{empty}

\begin{abstract}
In a previous article~\cite{Conde:2016csj}, we have argued that Low's sub-leading soft photon theorem can be recovered as a Ward identity associated to the same large gauge transformations that control the leading piece of the theorem. The key for that was to link the energy expansion displayed in the soft theorem to a $\frac{1}{r}$ expansion that we can perform in the associated asymptotic charge. We expect this idea to be valid in general, and here we provide compelling evidence for it by showing how the same method works in the case of Einstein-Hilbert gravity. More precisely, we are able to derive the three orders of the tree-level soft graviton theorem simply from the BMS supertranslation charge, known to give rise to the leading soft graviton theorem. In particular, we do not need to invoke superrotations (nor extended superrotations) at any point of the argument.
\end{abstract}

\cleardoublepage

\tableofcontents

\section{Introduction}

In the last few years there has been a renewed interest on soft theorems. Although the latter were first investigated long time ago~\cite{Low:1954kd,GellMann:1954kc,Low:1958sn,Weinberg:1965nx,Burnett:1967km,Bell:1969yw} and they are well established in phenomenological applications, the new enthusiasm comes from a more theoretical side. While from this second perspective there are many reasons to like soft theorems,\footnote{
For instance one can classify and reconstruct all tree-level scattering amplitudes of a large class of scalar theories~\cite{Cheung:2014dqa,Cheung:2015ota,Cheung:2016drk}, or more generically establish various constraints on effective actions of theories displaying spontaneous symmetry breaking~\cite{Bianchi:2016viy}. More conjecturally, particular soft behaviors of a theory may be related to its scattering amplitudes having a Cachazo-He-Yuan representation~\cite{Cachazo:2013hca,Cachazo:2016njl,Freddy}.}
arguably the jumpstarting source of fuel for the recent activity can be considered the connection between soft theorems and symmetries at null infinity, proposed by Strominger in 2013~\cite{Strominger:2013lka,Strominger:2013jfa}. Since then, this connection has accumulated considerable evidence in its favor, as we briefly overview below.

Historically, the two best studied examples of soft theorems are found in the four-dimensional processes where a soft photon or a soft graviton, with momentum $q\to0$, is emitted in a scattering process. Such a process can involve any other particle, as long as it is minimally coupled (otherwise the original soft theorems must be slightly modified~\cite{Elvang:2016qvq}) to the photons or gravitons. Soft theorems can then be thought as factorization properties that scattering amplitudes must obey in a low-energy expansion:
\begin{equation}
\label{softh}
	M_{n+1}\big(p_1,\ldots,p_n,q\big)=\left(\frac{S^{(0)}}{E_q}+\cdots+E_q^{s-1}S^{(s)}\right)M_{n}(p_1,\ldots,p_n)+\cO{E_q^{s}} \,,
\end{equation}
where $E_q$ and $s$ are the energy and spin of the emitted boson. For $s=1,2$, soft theorems display several orders in the energy $E_q$ of the emitted boson. Since actually $M_n$ only has support in $E_q=0$, a convenient way of rewriting the different orders in~\eqref{softh} is as
\begin{equation}
\label{subsofths}
	\lim_{E_q\to0}\partial_{E_q}^kE_q\,M_{n+1}=k!\,S^{(k)}M_n\,.
\end{equation}
It is customary to talk about several soft theorems, one per each order $k=0,\ldots,s\,$.

Regarding the leading order, both the soft photon and graviton theorems start at $\cO{E_q^{-1}}$, a reflection that their interactions are long-ranged. At this order, soft factorization is known to be non-renormalized~\cite{Weinberg:1965nx}, suggesting the existence of some underlying symmetry preserved at quantum level. Along this line, what Strominger proposed in~\cite{Strominger:2013lka,Strominger:2013jfa} was that these soft theorems could be rewritten as Ward identities of certain asymptotic (\textit{broken}) symmetries. More particularly, a class of residual gauge transformations for Maxwell theory would be behind the soft photon theorem, while in the case of Einstein-Hilbert gravity the constraints on emission of soft gravitons would come from the so-called \textit{supertranslations}, which form the Abelian ideal part of the Bondi-Metzner-Sachs (BMS) transformations~\cite{Bondi:1962px,Sachs:1962wk,Sachs:1962zza}. This proposal was notably checked to be true in subsequent works, namely~\cite{He:2014cra} (see also~\cite{Mohd:2014oja,Campiglia:2015qka,Kapec:2015ena}) for soft photons, and~\cite{He:2014laa} (see also~\cite{Campiglia:2015kxa}) for soft gravitons. Let us mention here that a successful connection also exists for the case of soft gluons~\cite{He:2015zea}, which from the perspective of scattering amplitudes behave similarly to soft photons, although their non-Abelian nature makes their story richer. Historically this case has attracted less attention because of the infrared nature of Yang-Mills theory, where free gluons cannot be observed.

Regarding the sub-leading orders, in our opinion the status of the connection has not been completely settled. Staying at tree level (we comment on loop level below), the soft graviton theorem stretches to sub-sub-leading order~\cite{Cachazo:2014fwa} (see~\cite{Gross:1968in,Jackiw:1968zza,White:2011yy} for earlier, more restricted, versions). As originally motivated by~\cite{Cachazo:2014fwa}, there have been several attempts~\cite{Kapec:2014opa,Campiglia:2014yka,Campiglia:2015yka} to link the sub-leading order to an extension of the BMS group proposed by Barnich and Troessaert~\cite{Barnich:2009se,Barnich:2010eb,Barnich:2011ct,Barnich:2011mi}, who enlarged it with the so-called \textit{superrotations}, a class of asymptotically-flat-preserving diffeomorphisms that extend the quotient part of BMS/supertranslations. But actually, the asymptotic symmetries which are used in~\cite{Kapec:2014opa} (and later in~\cite{Campiglia:2014yka,Campiglia:2015yka}) are not superrotations,\footnote{
Earlier in~\cite{Adamo:2014yya,Geyer:2014lca} the sub-leading soft graviton theorem was recovered from twistor-string-like constructions, but the connection to the superrotations of~\cite{Barnich:2009se,Barnich:2010eb,Barnich:2011ct,Barnich:2011mi} is not completely clear to us.
}
but a sort of generalization thereof, which actually do not preserve the asymptotic flatness condition. The authors in~\cite{Campiglia:2016jdj,Campiglia:2016efb} have recently proposed another set of asymptotic transformations presumably responsible for the sub-sub-leading soft graviton theorem, but some aspects of this proposal are unclear, among them again the issue of the transformations violating asymptotic flatness.

\smallskip

In this paper, we propose a consistent picture for both sub- and sub-sub-leading soft graviton theorems following from supertranslations, in particular with no need of invoking superrotations, nor the extended superrotations just mentioned above. The picture we want to advocate is one that we have recently used for Low's sub-leading soft photon theorem~\cite{Conde:2016csj}.\footnote{
We note here that, using ideas similar to those presented in~\cite{Campiglia:2016jdj,Campiglia:2016efb}, the authors there also have a proposal regarding the sub-leading soft photon theorem~\cite{Campiglia:2016hvg}, suggesting that it should be rewritten as the Ward identity of certain divergent large gauge transformations (namely proportional to $r$). Apart from the issue of the divergence, the nature of such transformations is uncertain since they break the required gauge condition.
}
The intuitive idea is the following: sub-leading orders of the soft factorization in the energy $E_q$ should correspond, in the bulk, to exploring the structure of null infinity at sub-leading orders in the inverse radial coordinate, $\frac{1}{r}$. Let us be a bit more explicit.

Assume that one can rewrite a given soft theorem as
\begin{equation}
\label{Ward}
	\langle\rm{out}|Q^{\textrm{out}}-Q^{\textrm{in}}|\rm{in}\rangle=0 \quad\Leftrightarrow\quad
	\langle\rm{out}|Q_{\rm{NL}}^{\textrm{out}}-Q^{\textrm{in}}_{\rm{NL}}|\rm{in}\rangle=
	-\langle\rm{out}|Q^{\textrm{out}}_{\rm{L}}-Q^{\textrm{in}}_{\rm{L}}|\rm{in}\rangle\,,
\end{equation}
where the charge $Q$ is ``broken'', and therefore can be decomposed into linear and non-linear pieces: $Q=Q_{\rm{L}}+Q_{\rm{NL}}$. The non-linear piece creates a zero-frequency boson when acting on the vacuum, so that the equation on the right of~\eqref{Ward} becomes the soft theorem~\eqref{softh}. If the charge $Q$ can be associated to an asymptotic symmetry, classically we will be able to write it as an integral at null infinity of a certain combination of physical fields,
\begin{equation}
	Q=\int_{\skyp}\td z\td\bz\td u\,C(\varphi_{\textrm{phys}})\,.
\end{equation}
Studying the classical phase space of the theory, we can solve the equations of motion in a $\frac{1}{r}$ expansion once boundary data has been specified. Thus physical fields can be expanded in such a way, and we can also formally expand the charge as
\begin{equation}
\label{Q012}
	Q=Q^{(0)}+\frac{Q^{(1)}}{r}+\frac{Q^{(2)}}{r^2}+\cdots\,.
\end{equation}
Then, we can convert the Ward identity~\eqref{Ward} into several ones, one at each $\frac{1}{r}$ order. Actually, since we want to evaluate the integrals at $r=\infty$, for the sub-leading orders it is better to use the expression
\begin{equation}
\label{Qn}
	Q^{(n)}=\frac{1}{n!}\int_{\skyp}\td z\td\bz\td u\,\lim_{\frac{1}{r}\to0}\left[\frac{\partial^{n}}{\partial (\frac{1}{r})^n}C(\varphi_{\textrm{phys}})\right]\,.
\end{equation}
We can then better notice the resemblance of $\langle\rm{out}|Q^{(n),\textrm{out}}-Q^{(n),\textrm{in}}|\rm{in}\rangle=0$ with the form of sub-leading soft theorems in~\eqref{subsofths}, which involve taking derivatives in $E_q$. In this way, the $E_q$ expansion of the soft theorem~\eqref{softh} should be matched with the $\frac{1}{r}$ expansion of the charge~\eqref{Q012}.

But there is a catch in this game. Although the expansion in~\eqref{Q012} contains an infinity of orders and we would seemingly get an infinite number of sub-leading soft theorems, only a finite number of these makes sense. There is only a finite number of \textit{boundary} fields, defined as the leading (in $\frac{1}{r}$) orders of the field-strength and Weyl tensor in gauge theory and gravity respectively. We use the prescription that only these boundary fields ``live at infinity'', therefore the sub-leading Ward identities are physical only when taking the derivatives in~\eqref{Qn} generate just boundary fields. The number of times this happens depends on the theory, and it precisely turns out that according to this criterium we can take one derivative in gauge theory and two for gravity. We will later give a less hand-wavy argument for why this is the case.

\smallskip

The strategy described above has been proven successful for showing how the sub-leading soft photon theorem follows from the leading one~\cite{Conde:2016csj}. In what follows we give further supporting evidence for this idea by showing how it also perfectly works in the case of the soft graviton theorem, which displays two sub-leading orders.

To be more precise, let us clarify that we do not work out the most general instance of the soft graviton theorem. Although the idea that we present should be valid for any setup, we focus on the particular case of scalar massless matter coupled to linearized gravity. By linearized gravity we mean, in terms of Feynman diagrams, that we only consider vertices where a graviton couples to two scalars. The graviton three-point vertex is left out in order to simplify the form of the stress-energy tensor, which otherwise would contain many non-linear terms coming from the graviton self-interaction. From a spectator at null infinity, massless matter is just simpler to deal with than massive matter, and the generalization to massive matter has already been considered in the literature~\cite{Campiglia:2015kxa}. Whether the matter is scalar or has spin would only change the form of the corresponding stress-energy tensor and make the action of the soft factors on amplitudes a bit more involved, which would just add technical difficulty.

Since our framework is based on a classical analysis of the equations of motion, we expect to reproduce just tree-level results. However we will see that, even classically, at sub-sub-leading order there is an ambiguity that we cannot fix, namely certain terms that can appear in the $Q^{(2)}$ of~\eqref{Q012}. At loop level, if one neglects IR corrections connected to collinear singularities~\cite{Naculich:2011ry,Akhoury:2011kq}, the loop corrections to the soft graviton theorem are expected only at sub-sub-leading order~\cite{Bern:2014oka,He:2014bga,Bern:2014vva}. We believe that this is not a coincidence, as we will comment in the final part of the manuscript.

\smallskip

The plan of the paper is as follows. In Section~\ref{linearGR} we discuss in detail the classical phase space of linearized general relativity, solving the equations of motion in a $\frac{1}{r}$ expansion. In that section we derive the expansion~\eqref{Q012} for the supertranslation charge that we later use in Section~\ref{sec:amps} to match with the soft graviton theorem. After discussing some of the implications and future directions of our work, we complete the article with three appendices which detail some passages of the text.

\section{Linearized gravity theory}
\label{linearGR}

A crucial step to recast symmetries at null infinity as S-matrix relations is the identification of symmetries at future null infinity $\skyp$ with symmetries at past null infinity $\skym$ \cite{Strominger:2013jfa}. Such identification was shown to be allowed only in the so-called Christodoulou-Kleinerman spaces \cite{Christodoulou:1993uv}, as for instance a finite neighborhood of Minkowski space. This issue is related to the stability of spacetimes. Though the study of symmetries at both null infinities is very well developed, whether we can obtain from them a symmetry of the S-matrix in the full Einstein gravity theory is still debatable. In contrast, the linearized gravitational theory on a fixed spacetime background, \textit{i.e.} Minkowski space, provides a well defined system that allows us to explore the connection between soft graviton theorems and symmetries at null infinity without any subtlety. We will illustrate our main results only at future null infinity throughout this paper; the counterpart of past null infinity and the identification can be given easily following~\cite{Strominger:2013jfa}.

To work with physical fields at future null infinity $\skyp$, it is usually convenient to adopt retarded coordinates:
\be\label{metric}
ds^2=\eta_{\mu\nu}\td x^\mu \td x^\nu=-\td u^2-2\td u\td r+2r^2\ga\, \td z\td \bz\,,
\ee
where $\ga=\frac{2}{(1+z\bz)^2}$. This will be the background spacetime we are about to linearize the full Einstein theory around: $g_{\mu\nu}=\eta_{\mu\nu}+h_{\mu\nu}$. The non-zero connection coefficients associated to \eqref{metric} are
\be
\Gamma^u_{z\bz}=r \ga\,,\quad
\Gamma^r_{z\bz}=-r\ga\,,\quad
\Gamma^z_{rz}=\frac{1}{r}\,,\quad
\Gamma^z_{zz}=\p_z\ln\ga\,,\quad
\Gamma^{\bz}_{\bz\bz}=\p_{\bz}\ln\ga\,.
\ee
Consequently, the Pauli-Fierz equations should be organized in a covariant way:
\be
\label{Einsteineqs}
E_{\mu\nu}\equiv\cd_\mu\cd_\nu h + \cd^\tau\cd_\tau h_{\mu\nu} - \cd^\tau\cd_\mu h_{\nu\tau} - \cd^\tau \cd_\nu h_{\mu\tau} - \eta_{\mu\nu}(\cd^\tau\cd_\tau h-\cd^\tau\cd^\rho h_{\tau\rho})=T_{\mu\nu}\,,
\ee
where $T_{\mu\nu}$ is a generic conserved stress-energy tensor and $h=\eta^{\mu\nu}h_{\mu\nu}$. All the indices should be lowered and raised with $\eta$. Notice that we are using natural units where $8\pi G_{\textrm{N}}=1\,$.

We will work in the Newman-Unti gauge \cite{Newman:1962cia}, for which
\be
h_{rr}=h_{rz}=h_{r\bz}=h_{ru}=0\,.\label{condition}
\ee
We believe that this choice is more convenient than the Bondi gauge in practice. In the full Einstein theory, the Newman-Unti gauge is connected to the Bondi gauge by a radial transformation~\cite{Barnich:2011ty}. Hence, the gauge condition we are using here is definitely equivalent to the one adapted from Bondi gauge in \cite{Hawking:2016sgy} recently. Those two ans\"atze are connected by a trivial gauge transformation.
To preserve asymptotic flatness, one needs to require the following asymptotic behaviors for the linearized fields:
\be
h_{uu}=\cO{r^{-1}}\,,\;\;h_{uz}=\cO1\,,\;\;h_{u\bz}=\cO1\,,\;\;h_{zz}=\cO r\,,\;\;h_{\bz\bz}=\cO r\,,\;\;h_{z\bz}=\cO1\,.
\ee


\subsection{Asymptotic symmetries}

Let us first discuss what are the asymptotic symmetries of the linearized gravity theory. We know that this theory is invariant under the gauge transformation $\delta h_{\mu\nu}=\nabla_\mu \epsilon_\nu + \nabla_\nu \epsilon_\mu$. We are interested in the residual gauge transformations that preserve the conditions
\begin{gather}
\delta_\epsilon h_{rr}=\delta_\epsilon h_{rz}=\delta_\epsilon h_{r\bz}=\delta_\epsilon h_{ru}=0\,,\label{gauge}\\
\begin{aligned}
\delta_\epsilon h_{uz}&=\cO1\,, & \delta_\epsilon h_{u\bz}&=\cO1\,, & \delta_\epsilon h_{zz}&=\cO r\,,\\
\delta_\epsilon h_{\bz\bz}&=\cO r\,, & \delta_\epsilon h_{z\bz}&=\cO1\,, & \delta_\epsilon h_{uu}&=\cO {r^{-1}}\,.\label{boundary}
\end{aligned}
\end{gather}
The general solution of $\epsilon$ to the gauge condition \eqref{gauge} is
\begin{gather}
  \left\{\begin{array}{l}
      \epsilon_r=-f(u,z,\bz)\,,\\
\epsilon_z=r^2\bY(u,z,\bz)-r\,\p_z f\,,\\
\epsilon_{\bz}=r^2Y(u,z,\bz)-r\,\p_{\bz} f\,,\\
\epsilon_u=r\,\p_u f - X(u,z,\bz)\,,
\end{array}
\right.
\end{gather}
where $f,X,Y$ are arbitrary functions.
The first and second equations of \eqref{boundary} imply that $\p_u \bY(u,z,\bz)=0=\p_u Y(u,z,\bz)$. The third and fourth ones require that $\p_z[\bY(z,\bz)\gai]=0$ and $\pb [Y(z,\bz)\gai]=0$. Hence, $\bY(z,\bz)=\bY(\bz)\ga$ and $Y(z,\bz)=Y(z)\ga$. The fifth one determines $X=f+\frac{1}{\ga}\p_z\p_{\bz}f$ and $\p_u f=\frac{1}{2\ga}[\p_z(Y\ga)+\p_{\bz}(\bY\ga)]$. The last one imposes no further constraint on $\epsilon$.
The residual gauge transformation can be finally written as
\bea
\label{rgt}
\epsilon_\mu=\left[r\,\p_u f - f - \gai\p_z\p_{\bz}f\,,\;-f\,,\;-r\,\p_z f + r^2\ga\,\bY(\bz)\,,\;-r\,\p_{\bz} f + r^2\ga\, Y(z)\right]\,,
\eea
where $f=f(z,\bz)+\frac{u}{2\ga}[\p_z(Y\ga)+\p_{\bz}(\bY\ga)]$. It may be more convenient to have the residual gauge transformation in vector form by raising the index with $\eta^{\mu\nu}$:
\be
\epsilon^\mu=\left[f\,,\;-r\,\p_u f + \frac{1}{\ga}\,\p_z\p_{\bz}f\,,\;Y-\frac{\p_{\bz}f}{r\ga}\,,\;\bY-\frac{\p_{z}f}{r\ga}\right]\,.
\ee
We have thus established the form of the residual gauge transformations of the linearized theory. The vectors $\epsilon^\mu$ are therefore asymptotic Killing vectors of the background Minkowski spacetimes. They are exactly the same as the BMS vectors found in \cite{Barnich:2010eb} when the solution metric is chosen as the Minkowski spacetimes $\eta_{\mu\nu}$.

The symmetry algebra is defined through $[\delta_{\epsilon_1},\delta_{\epsilon_2}]h_{\mu\nu}=\delta_{[\epsilon_1,\epsilon_2]_{_l}}h_{\mu\nu}$ . One can check easily that in the linearized case, the asymptotic symmetry algebra is not the standard Lie algebra between vectors, but an Abelian algebra such that $[\epsilon_1,\epsilon_2]_{_l}=0$ where the subscript ``$l$'' denotes the algebra in linearized theory. This is another effect of linearization.

\subsection{The solution space for linearized gravity theory coupled to generic matter fields}

In this section, we want to solve the equations of motion in a $\frac{1}{r}$ expansion around future null infinity. To do that, we start with the arrangement of the equations of motion \eqref{Einsteineqs}.
As we are dealing with a gauge theory, not all the equations of motion are independent. The constraints among them are inherited from the Bianchi identity of the full Einstein theory, which can be written as $\nabla_\mu (E^{\mu\nu}-T^{\mu\nu})=0$. Taking into account such constraints, we are able to arrange the ten equations of motion as follows:
\begin{multicols}{2}
\begin{itemize}
\item Four hypersurface equations:
\bea
E_{r\mu}=T_{r\mu}\,.
\eea
\item Two standard equations:
\bea
E_{zz}=T_{zz}\,,\;\;E_{\bz\bz}=T_{\bz\bz}\,.
\eea
\item One trivial equation:
\bea
E_{z\bz}=T_{z \bz}\,.
\eea
\item Three supplementary equations:
\bea
E_{uz}=T_{uz}\,,\;E_{u\bz}=T_{u\bz}\,,\;E_{uu}=T_{uu}\,.
\eea
\end{itemize}
\end{multicols}
\noindent
The advantage of such arrangement is well explained in the literature \cite{Bondi:1962px,Sachs:1962wk,Barnich:2010eb}. We will set all the radial components of the stress-energy tensor $T_{r\mu}$ to zero to adapt to our gauge condition \eqref{condition}. This can be done by using the ambiguities of a conserved stress-energy tensor. The details are given in Appendix \ref{solutionspace}. To preserve asymptotic flatness, the remaining components of the stress-energy tensor should have the following asymptotic behaviors:
\be\begin{aligned}
T_{uu}&=\cO{r^{-2}}\,, & T_{uz}&=\cO{r^{-2}}\,, & T_{u\bz}&=\cO{r^{-2}}\,,\\
T_{zz}&=\cO{r^{-2}}\,, & T_{\bz\bz}&=\cO{r^{-2}}\,, & T_{z\bz}&=\cO{r^{-2}}\,.
\end{aligned}\ee
Now we are ready to solve out the solution space of the linearized theory. We begin with the hypersurface equation $E_{rr}=0$.
We find that
\be
E_{rr}=\frac{2\p^2_r(\frac{h_{z\bz}}{r})}{r\ga}\,.
\ee
Because of the boundary condition $h_{z\bz}=\cO{1}$, this will yield $h_{z\bz}=0$ . The $rz$-component of the equations of motion can be rewritten as
\be
E_{rz}=\frac{1}{r^2}\p_r\left[r^4\p_r\left(\frac{h_{uz}}{r^2}\right)\right]-\frac{1}{\ga}\p_{\bz}\p_r\left(\frac{h_{zz}}{r^2}\right)+\p_z\p_r\left(\frac{h_{z\bz}}{r^2\ga}\right)\,.
\ee
Once $h_{z\bz}=0$ has been imposed, from $E_{rz}=0$ we get
\bea
\label{AM}
h_{uz}=\frac{2\sqrt{\ga}\,\xbar\Psi_1^0(u,z,\bz)}{3r} + \frac{r^2}{\ga}\;\pzb\;\int^{\infty}_r\; \frac{\td r'}{{r'}^4}\;\int^{\infty}_{r'}\td r''\;{r''}^2\p_{r''}\left(\frac{h_{zz}}{{r''}^2}\right)\,,
\eea
where $2\sqrt{\ga}\,\xbar\Psi_1^0(u,z,\bz)$ is an integration constant respect to $r$, and the notation has been chosen to make contact with the result of Newman and Penrose~\cite{Newman:1968uj}. To avoid logarithmic terms, one has to accept a gap in the expansion of $h_{zz}$, namely $h_{zz}=\cO r+\cO {r^{-1}}$. Logarithmic terms would break the expansion in $\frac{1}{r}$ and lead to a new type of expansion which is called polyhomogeneous expansion \cite{Chrusciel:1993hx}, involving both an expansion in $\frac{1}{r}$ and in $\ln r$. On the other hand, it is amusing to see that the gauge transformation~\eqref{rgt} does not affect the $\cO1$ order of $h_{zz}$. Hence, this gap does not give further constraints on the asymptotic symmetry vectors $\epsilon^\mu$.

Simply swapping $z$ by $\bz$, from $E_{r\bz}=0$ one can get
\bea\label{AMb}
h_{u\bz}=\frac{2\sqrt{\ga}\,\Psi_1^0(u,z,\bz)}{3r} + \frac{r^2}{\ga}\;\p_z\;\int^{\infty}_{r}\; \frac{\td r'}{{r'}^4}\;\int^{\infty}_{r'}\;\td r''\;{r''}^2\p_{r''}\left(\frac{h_{\bz\bz}}{{r''}^2}\right)\,,
\eea
We continue with $E_{ru}=0$, which yields
\bea
E_{ru}&=&-\frac{2}{r^2}\p_r\left(r h_{uu}\right) + \frac{1}{r^4\ga}\left[\p_r\left(r^2\p_{\bz}h_{uz}\right)+\p_r\left(r^2\p_z h_{u\bz}\right)\right] \nn\\&&- \frac{1}{r^4\ga}\left[\p_{\bz}\left(\frac{\p_{\bz}h_{zz}}{\ga}\right) + \p_z\left(\frac{\p_z h_{\bz\bz}}{\ga}\right)\right]\nn\\
&&+\frac{2}{r^4\ga}\left[h_{z\bz} + \p_z\p_{\bz}\left(\frac{h_{z\bz}}{\ga}\right)\right] + \frac{2}{r\ga}\p_r\left(\frac{\p_r h_{z\bz}}{r}\right) - \frac{2}{r^2\ga}\p_u\p_r h_{z\bz}.
\eea
By inserting the solution of $h_{z\bz}$ and $h_{uz}$, one obtains
\bea\label{mass}
h_{uu}=\frac{M(u,z,\bz)}{r}&-&\frac{1}{2r\ga}\;\pzb\;\int^{\infty}_{r}\;\frac{\td r'}{{r'}^2}\;\left[\p_{r'}\left({r'}^2 h_{uz}\right)-\frac{\p_{\bz}h_{zz}}{\ga}\right]\nn\\
&-&\frac{1}{2r\ga}\;\pz\;\int^{\infty}_{r}\;\frac{\td r'}{{r'}^2}\;\left[\p_{r'}\left({r'}^2 h_{u\bz}\right)-\frac{\p_{z}h_{\bz\bz}}{\ga}\right],
\eea
where $M(u,z,\bz)$ is another integration constant.

When all the four hypersurface equations are satisfied, the trivial equation will be satisfied automatically. We now turn to the standard equations. Suppose that $h_{zz}$, $h_{\bz\bz}$ and the non-vanishing components of the stress-energy tensor are given as initial data in the following way:
\bea
&&h_{zz}=2\ga \xbar\sigma^0(u,z,\bz)r+\frac{\bar{\Psi}^0_0(u,z,\bz)\ga}{3r}+\sum\limits_{m=1}^\infty\frac{\bar{\Psi}_0^{(m)}(u,z,\bz)\ga}{r^{m+1}}\,,\\
&&h_{\bz\bz}=2\ga \sigma^0(u,z,\bz)r+\frac{{\Psi}^0_0(u,z,\bz)\ga}{3r}+\sum\limits_{m=1}^\infty\frac{\Psi_0^{(m)}(u,z,\bz)\ga}{r^{m+1}}\,,\\
&&T_{\mu\nu}=\frac{T_{\mu\nu}^0(u,z,\bz)}{r^2}+\sum\limits_{m=1}^\infty\frac{T_{\mu\nu}^{(m)}(u,z,\bz)}{r^{m+2}}\,.
\eea
Inserting these expansions into \eqref{AM}-\eqref{mass}, we arrive at
\begin{align}
\label{huz}
&h_{uz}=\sqrt{\ga}\,\eth\xbar\sigma^0 + \frac{2\sqrt{\ga}\xbar\Psi_1^0}{3r} - \frac{\sqrt{\ga}\eth\xbar\Psi_0^0}{4r^2}-\sum\limits_{m=1}^\infty\frac{(m+3)\sga}{(m+1)(m+4)}\frac{\eth\xbar\Psi_0^{(m)}}{r^{m+2}}\,,\\
&h_{u\bz}=\sqrt{\ga}\,\xbar\eth\sigma^0 + \frac{2\sqrt{\ga}\Psi_1^0}{3r} - \frac{\sqrt{\ga}\xbar\eth\Psi_0^0}{4r^2}-\sum\limits_{m=1}^\infty\frac{(m+3)\sga}{(m+1)(m+4)}\frac{\xbar\eth{\Psi}_0^{(m)}}{r^{m+2}}\,,\label{huzb}\\
&h_{uu}=\frac{M}{r}-\frac{\eth\xbar\Psi_1^0+\xbar\eth\Psi_1^0}{3r^2} + \frac{\eth^2\xbar\Psi_0^0+\xbar\eth^2\Psi_0^0}{12r^3}+\sum\limits_{m=1}^\infty\frac{1}{(m+1)(m+4)}\frac{\eth^2\xbar\Psi_0^{(m)}+\xbar\eth^2{\Psi}_0^{(m)}}{r^{m+3}}\,,\label{huu}
\end{align}
where $\eth$ and $\xbar\eth$ are derivative operators originally introduced in~\cite{Newman:1966ub} to replace the covariant derivatives on a sphere. We believe they are a more convenient choice for the calculation on null infinity, which has topology $S^2\times \mathbb{R}$. The definition plus some useful properties of $\eth$ and $\xbar\eth$ are listed in Appendix \ref{eth}.

The standard equation $E_{zz}=T_{zz}$, that is recast as
\be\label{standard}\begin{split}
-2r\p_u\p_r\frac{h_{zz}}{r}+2\ga\p_z\frac{\p_r h_{uz}}{\ga}+r\p_r^2\frac{h_{zz}}{r}=T_{zz}\,,
\end{split}\ee
controls the time evolution of the initial data $h_{zz}$, except the leading order $\xbar\sigma^0$. Following the terminology of \cite{Bondi:1962px}, we refer to $\dot{\xbar\sigma}^0$ as a \textit{news} function. There is another \textit{news} function, $\dot{\sigma}^0$, from the counterpart in $E_{\bz\bz}=T_{\bz\bz}$. The precise information that is extracted from~\eqref{standard} is:
\begin{align}
&\p_u\Psi_0^0=\eth\Psi_1^0+\frac{3}{4\ga}T_{\bz\bz}^0\,,\label{Phi0}\\
&\p_u\Psi^1_0=\frac{T_{\bz\bz}^1}{3\ga}-\frac13(\eth\xbar\eth\Psi_0^0+2\Psi_0^0)\,,\\
&\p_u\Psi^{m+1}_0=\frac{T_{\bz\bz}^{m+1}}{(m+3)\ga}-\frac{2(m+2)}{(m+1)(m+4)}\left(\eth\xbar\eth\Psi_0^m+\frac{(m+1)(m+4)}{2}\Psi_0^m\right)\,.
\end{align}
The supplementary equations are much simplified once all the hypersurface equations and standard equations are satisfied, due to the existence of the Bianchi identity. As is well explained in \cite{Bondi:1962px,Sachs:1962wk,Barnich:2010eb}, only one order in the expansion of the supplementary equations will be left to solve. At last, the supplementary equations fix the time evolution of the integration constants as
\bea
&&\p_u M=\eth^2\dot{\xbar\sigma}^0+\xbar\eth^2\dot{\sigma}^0+\half T_{uu}^0\,,\\
&&\p_u\Psi_1^0=\half\eth M+\half(\eth^3\xbar\sigma^0-\eth\xbar\eth^2\sigma^0)+\frac{1}{2\sga}T_{u\bz}^0\,,
\eea
We can reorganize the time evolution equations by solving $M=M_0(z,\bz)+\eth^2{\xbar\sigma}^0+\xbar\eth^2{\sigma}^0+\half\int \td u\;T^M_{uu}$. Define $\Psi_2^0=\eth^2{\xbar\sigma}^0+\half(M_0 +\half\int \td u\;T^M_{uu})$. Hence
\bea
&&M=\Psi_2^0+\xbar\Psi_2^0\,,\nn\\
&&\p_u \Psi_2^0=\eth^2\dot{\xbar\sigma}^0+\frac{1}{4} T^0_{uu}\,,\label{Phi2}\\
&&\p_u \Psi_1^0=\eth\Psi_2^0 + \frac{1}{2\sga} T^0_{u\bz}\,.\label{Phi1}
\eea

The action of the asymptotic symmetries on the solution space is quite simple in the
linearized case:
\begin{align}
&\delta\sigma^0=-\eth^2f(z,\bz)-\frac{u}{2}\eth^3\cY\,,\quad\delta\xbar\sigma^0=-\xbar\eth^2f(z,\bz)-\frac{u}{2}\xbar\eth^3\xbar\cY\,,\\
&\delta\Psi_a^0=0\,,\,a=0,1,2\,,\quad\delta\Psi_0^m=0\,,\,m=1,2,\cdot\cdot\cdot\,,\\
&\cY=\bY(\bz)\sqrt{\ga}\,,\quad\bar\cY=Y(z)\sqrt{\ga}\,,\nn
\end{align}
where $f(z,\bz)$ defines the supertranslation and $\cY$, $\bar\cY$ define superrotation transformation on the phase space respectively. This can be understood easily from a physical interpretation. In Newman-Penrose formalism, all the $\Psi$s in the solution space are components of the Weyl tensor. They are the analogue of the field strength in electromagnetism which are invariant under gauge transformations. Hence \eqref{huu} yields\be\delta h_{uu}=0\,.\ee

\subsection{The asymptotic conserved charges}
\label{ssec:asympQ}

Although associating conserved charges to asymptotic symmetries is a well-established issue \cite{Wald:1999wa,Barnich:2001jy}, how to extend those charges to the next-to-leading orders is quite a tricky question \cite{Glenn}. In practice, Newman and Penrose have demonstrated in \cite{Newman:1968uj} how to construct conserved quantities at each order of a series expansion in $\frac1r$. A very short review of the relevant results is given in Appendix \ref{NP}.
Remarkably, in our metric formalism, all the conserved quantities constructed by Newman and Penrose can be unified in a very simple and elegant form which is given by
\be\label{charge}
Q=\int \td z \td\bz\,\ga\;r\,f(z,\bz)\,h_{uu}\,.
\ee
The charge can be defined at any cross section of $\skyp$. Inserting the $\frac{1}{r}$ expansion of $h_{uu}$, written in~\eqref{huu}, into this charge gives an expansion like~\eqref{Q012}:
\begin{multline}\label{charge'}
Q=\int \td z \td\bz\,\ga\;f(z,\bz)\,\bigg[\Psi_2^0+\xbar\Psi_2^0-\frac{\eth\xbar\Psi_1^0+\xbar\eth\Psi_1^0}{3r} + \frac{\eth^2\xbar\Psi_0^0+\xbar\eth^2\Psi_0^0}{12r^2}\\
+\sum\limits_{m=1}^\infty\frac{1}{(m+1)(m+4)}\frac{\eth^2\xbar\Psi_0^{(m)}+\xbar\eth^2{\Psi}_0^{(m)}}{r^{m+2}}\bigg]\,.
\end{multline}
To recover the Newman-Penrose conserved quantities, one just need to identify $f(z,\bz)$ with certain spin-$s$ spherical harmonics (given in Appendix \ref{eth}). More precisely, putting for example $f(z,\bz)=\Preind_{0}{Y}_{0,0}$ yields
\be\label{Qc0}
Q^{(0)}=\int \td z \td\bz\,\ga\;\Preind_{0}{Y}_{0,0}\,(\Psi_2^0+\xbar\Psi_2^0),
\ee
which is the Bondi mass \eqref{masslaw}, and
\bea
&&Q^{(1)}=-\frac13\int \td z \td\bz\,\ga\;\Preind_{0}{Y}_{0,0}\,\left(\eth\xbar\Psi_1^0+\xbar\eth\Psi_1^0\right),\\
&&Q^{(2)}=\frac{1}{12}\int \td z \td\bz\,\ga\;\Preind_{0}{Y}_{0,0}\,\left(\eth^2\xbar\Psi_0^0+\xbar\eth^2\Psi_0^0\right),
\eea
which are the two identically vanishing quantities \eqref{trivial1} and \eqref{trivial2}. Instead $f(z,\bz)=\Preind_{0}{Y}_{m+1,l}$ yields
\be
Q^{(m)}=-\int \td z \td\bz\,\ga\;\Preind_{0}{Y}_{m+1,l}\,(\frac{\eth^2\xbar\Psi_0^{(m)}+\xbar\eth^2{\Psi}_0^{(m)}}{r^{m+2}})\,,\quad(m>2)\,,
\ee
which correspond to the new conservation laws \eqref{newlaws}. Indeed to recover \eqref{newlaws}, one just needs to perform an integration by parts twice.

A common observation of asymptotic conserved charges is that they should be some generalizations of exact conserved charges. Indeed, the supertranslation charge
\be\label{st}
Q_{\textrm{st}}=\int \td z \td\bz\,\ga\;f(z,\bz)\,\left(\Psi_2^0+\xbar\Psi_2^0\right)\,,
\ee
explored in \cite{Barnich:2011mi,Barnich:2013axa,He:2014laa} is a generalization of the Bondi mass \eqref{Qc0}. It is amusing to see that the supertranslation charge \eqref{st} is just the leading term of \eqref{charge}. Hence it is reasonable to propose that \eqref{charge} also gives the next-to-leading orders of the asymptotic conserved charges.

The Poisson bracket of the surface charges can be extended to next-to-leading orders now. It is obvious that the charge algebra is Abelian
\be
\{Q_{\textrm{st}},Q_{\textrm{st}'}\}=0\,,
\ee
since $\delta h_{uu}=0$. This should not be surprising in a linearized theory. Using the Poisson bracket, one gets the action of the charge on the phase space of radiative modes
\be
\{Q_{\textrm{st}},\sigma^0\}=-\eth^2f(z,\bz),\quad\{Q_{\textrm{st}},\xbar\sigma^0\}=-\xbar\eth^2f(z,\bz)\,.
\ee

Let us make manifest the first three orders of the charge~\eqref{charge} at ${\skyp_-}$, needed for recovering the soft graviton theorem in the next section. To that end, we insert the time evolution constraints \eqref{Phi2}, \eqref{Phi1} and \eqref{Phi0} as well as
the modified stress-energy tensor of complex scalar fields \eqref{modify} as follows:

\begin{itemize}
\item Leading charge:
\begin{equation}
\label{Qleading}
\begin{aligned}
Q^{(0)}&=\int_{\skyp_-}\;\td z\td\bz\,\ga \,f\left(\xbar\Psi_2^0+\Psi_2^0\right)\\
&=\int_{\skyp}\;\td z\td\bz \td u\,\ga \,f\left(\eth^2\dot{\xbar\sigma}^0+\xbar\eth^2\dot{\sigma}^0+\frac{1}{2}(\p_u\phi)\,(\p_u\xbar\phi)\right)\,.
\end{aligned}
\end{equation}

\item Sub-leading charge:
\begin{equation}
\label{Qsub}
\begin{aligned}
  -Q^{(1)}&=\frac13\int_{\skyp_-}\td z\td\bz\,\ga\,\left(\eth f\, \xbar\Psi_1^0+\xbar\eth f\, \Psi_1^0\right)\\
  &=\frac16\int_{\skyp}\td z\td\bz \td u\, \ga\,\bigg[-\frac{1}{2\sga}\xbar\eth f \left(\phi\pzb \p_u\xbar\phi+\xbar\phi\p_u\pzb \phi\right)
  -\frac{1}{2\sga}\eth f\left(\phi\p_u\p_z\xbar\phi+\xbar\phi\p_u\p_z\phi\right)\\
  &\hspace{2cm}+u\,\eth\xbar\eth f\,(\p_u\phi)(\p_u\xbar\phi)+2u\,\eth\xbar\eth f\left(\eth^2\dot{\xbar\sigma}^0+\xbar\eth^2\dot{\sigma}^0\right)\bigg]\,.
\end{aligned}
\end{equation}

\item Sub-sub-leading:
\bea\label{Qsubsub}
Q^{(2)}&=&\frac{1}{12}\int_{\skyp_-}\td z\td\bz \,\left(\eth^2f\,\xbar\Psi_0^0+\xbar\eth^2f\,\Psi_0^0\right)\,,\nn\\
&=&\frac{1}{48}\int_{\skyp}\td z\td\bz\td u\,\bigg[3\gai\left(\eth^2 f \p_z\phi\,\p_z\xbar\phi + \xbar\eth^2 f \pzb\phi\,\pzb\xbar\phi\right)\nn\\
&&\hspace{1cm}+\frac{2u}{\sqrt{\ga}}\left(\xbar\eth\eth^2f\,\left(-\half\phi\p_u\p_z\xbar\phi-\half\xbar\phi\p_u\p_z\phi\right)
+\eth\xbar\eth^2f\, \left(-\half\phi\p_u \pzb\xbar\phi-\half\xbar\phi\p_u \pzb\phi\right)\right)\nn\\
&&\hspace{1cm}+u^2\,\left(\eth^2\xbar\eth^2f\, (\p_u\phi)(\p_u\xbar\phi)\right)
+2u^2\,\left(\eth^2\xbar\eth^2f \left(\eth^2\dot{\xbar\sigma}^0+\xbar\eth^2\dot{\sigma}^0\right)\right)\nn\\
&&\hspace{1cm}-\frac{2}{\sqrt{\ga}}\left(\xbar\eth\eth^2f\, A_{\bz}+\eth\xbar\eth^2f\, A_z\right)+2u^2 \eth^2\xbar\eth^2f\, A_r\,.
\eea
\end{itemize}
In order to derive these expressions we have integrated by parts in several occasions, assuming that $\lim_{|u|\to\infty}\phi(u,z,\bz)=0$ and $\lim_{|u|\to\infty}A_\mu(u,z,\bz)=0$. These are necessary conditions for the field configurations that revert to vacuum at $\skyp_{\pm}$.

In the sub-sub-leading charge~\eqref{Qsubsub} we see two arbitrary $r$-integration constants, $A_z$ and $A_r$, showing up. They appear naturally in the modified stress-energy tensor~\eqref{modify}, but miraculously they cancel out of the leading and sub-leading charge. We postpone commenting more on this curious fact to the next sections.

\section{Matching the soft theorem}
\label{sec:amps}

As announced in the introduction, in this section we show how to recover the three orders of the tree-level soft graviton theorem as Ward identities of supertranslations. In the previous section we have laid out the form of the leading, sub- and sub-sub-leading supertranslation charges, namely equations~\eqref{Qleading}-\eqref{Qsubsub}. As in~\cite{Conde:2016csj}, we want to plug these charges in~\eqref{Ward} and rewrite the ensuing expressions in the more familiar amplitude form of soft theorems. We proceed order by order.

\subsection{Leading order}

Let us start by stating the leading soft graviton theorem. Depending on the helicity of the emitted soft graviton we can say that we have two soft theorems:
\begin{equation}
\label{leadsofth}
M^{\pm}_{n+1}(p_1,\ldots,p_n,q)=\sum_{k=1}^n \frac{\left(p_k\cdot \epsilon^\pm(q)\right)^2}{p_k\cdot q}\,M_{n}(p_1,\ldots,p_n)+\cO{q^0} \,.
\end{equation}
In the numerator of the soft factor we could have written more generally $p_k^{\mu}p_k^{\nu}\epsilon^{\pm}_{\mu\nu}(q)$. Since the graviton is a symmetric and traceless representation, we can take the graviton polarization tensor to be the product of two photon polarization vectors: $\epsilon_{\mu\nu}^{\pm}(q)=\epsilon_{\mu}^{\pm}(q)\epsilon_{\nu}^{\pm}(q)$. Notice also that, opposite to the soft photon theorem, no coupling constant shows up. Because of the equivalence principle, the coupling constant is the same for all legs, and is set to 1 in our units.

The first step is to translate \eqref{leadsofth} into spherical retarded coordinates \eqref{metric}. Let us collect here all the ingredients needed for this process:
\begin{gather}
\label{qinw}
q_{\mu}=\frac{E_q}{1+w\bar{w}}\left(1+w\bar{w},w+\bar{w},i(\bar{w}-w),1-w\bar{w}\right)\,,\\
\label{pk}
p_{(k)\mu}=\frac{E_k}{1+w_k\bar{w}_k}\left(1+w_k\bar{w}_k,w_k+\bar{w}_k,i(\bar{w}_k-w_k),1-w_k\bar{w}_k\right)\,,\\
\epsilon_{\mu}^-(q)=\frac{1}{\sqrt{2}}\left(\bar{w},1,-i,-\bar{w}\right)\,,\qquad
\epsilon_{\mu}^+(q)=\frac{1}{\sqrt{2}}\left(w,1,i,-w\right)\,,\\
\label{einw}
\begin{aligned}
\epsilon_z^-(q)&=\sqrt{2}\,r\,\frac{\bz(\bar{w}-\bz)}{(1+z\bz)^2}\,,\qquad   &	\epsilon_z^+(q)&=\sqrt{2}\,r\,\frac{1+\bz w}{(1+z\bz)^2}\,,\\
\epsilon_{\bz}^-(q)&=\sqrt{2}\,r\,\frac{1+z\bar{w}}{(1+z\bz)^2}\,,\qquad   &	\epsilon_{\bz}^+(q)&=\sqrt{2}\,r\,\frac{z(w-z)}{(1+z\bz)^2} \,.
\end{aligned}
\end{gather}
We are assuming here that all the momenta appearing in \eqref{leadsofth} are massless. When $q=(1,0,0,1)$, the polarization tensors are $\epsilon^{\pm}=\frac{1}{\sqrt{2}}(0,1,\pm i,0)$. We can also notice that when $\vec{q}$ is aligned with $\vec{x}=r\left(\frac{z+\bz}{1+z\bz},i\,\frac{z-\bz}{1+z\bz},\frac{1-z\bz}{1+z\bz}\right)$, $\epsilon^-_z$ and $\epsilon^+_{\bz}$ vanish.

Writing now $M_{n}(p_1,\ldots,p_n)=\langle\textrm{out}|\textrm{in}\rangle$, and $M^{\pm}_{n+1}(p_1,\ldots,p_n,q)=\langle\textrm{out}|\mathfrak{a}_{\mp}(-q)|\textrm{in}\rangle$,\footnote{
We are using crossing symmetry $\langle\textrm{out}|\mathfrak{a}_{\mp}(-q)=(\mathfrak{a}^{\dagger}_{\pm}(q)|\textrm{out}\rangle)^\dagger$.
}
we can re-express \eqref{leadsofth} for a negative-helicity graviton as
\begin{equation}
\label{soft-}
\langle\textrm{out}|\lim_{E_q\to0}E_q\,\mathfrak{a}_+(-q)|\textrm{in}\rangle
=(1+w\bw)\sum_{k=1}^n \frac{E_k}{w-w_k}\frac{\bw_k-\bw}{1+w_k\bw_k}\,\langle\textrm{out}|\textrm{in}\rangle\,.
\end{equation}
With the benefit of knowing what we will find below, let us differentiate this result, times $\gaw$, with respect to $\bw$. We obtain
\begin{equation}
\label{Dsoft-}
  \p_{\bw}\left[\gaw\lim_{E_q\to0}
  \langle\textrm{out}|E_q\,\mathfrak{a}_+(-q)|\textrm{in}\rangle\right]=
  -\gaw\sum_{k=1}^n\frac{1+w\bw_k}{1+w_k\bw_k}\frac{E_k}{w-w_k}\lim_{E_q\to0}\langle\textrm{out}|\textrm{in}\rangle\,.
\end{equation}
This equation is exactly what we obtain when we plug~\eqref{Qleading} in the Ward identity~\eqref{Ward}. Let us show it explicitly for the \textit{out} charges (the \textit{in} charges can be worked out analogously by repeating all the construction at past null infinity):

In the first place we extract the linear and non-linear pieces out of~\eqref{Qleading}. We also unfold the $\eth$-derivatives in terms of partial derivatives, in order to later perform integrations by parts more easily. Also, since we are focusing only on the negative-helicity component of the soft graviton theorem it is enough to keep the ``holomorphic half'' of the charge. This gives us
\begin{align}
\label{Q0NL}
  Q^{(0)}_{\textrm{NL}}&=\int\td u\td z\td\bz\,f\,\p_u\,\p_{\bz}\,\gai\p_{\bz}\,\ga\xbar\sigma^0\,,\\
\label{Q0L}
  Q^{(0)}_{\textrm{L}}&=\frac14\int\td u\td z\td\bz\,\gamma_{z\bz}\,f(z,\bz)\,T^0_{uu}\,,
\end{align}
where we should understand that the derivatives act on everything to their right. In order to find out how the non-linear piece acts on asymptotic states we must express $\xbar\sigma^0$ in terms of harmonic oscillators. We can simply use the expression found in~\cite{He:2014laa}.\footnote{
Comparing equation (2.1) in~\cite{He:2014laa} with our $h_{zz}=2\ga \xbar\sigma^0(u,z,\bz)r+\frac{\bar{\Psi}^0_0(u,z,\bz)\ga}{3r}+O(r^{-2})$, we see that $C_{zz}=2\gamma_{z\bz}\bar\sigma^0$.}
In our units, we get
\begin{equation}
\label{bs0}
\xbar\sigma^0(u,z,\bz)=\frac{i}{8\pi^2}\int_0^\infty\td E_q\left[\mathfrak{a}_+(E_q\hat x)e^{-iE_q u}-\mathfrak{a}^\dagger_-(E_q\hat x)e^{iE_q u}\right]\,,
\end{equation}
where $\hat{x}$ is the unit vector pointing to the point $(z,\bz)$ on the sky. Because~\eqref{bs0} has the form of a Fourier transform, we recall the following relation that will be useful below. Defining the Fourier transform through $F(u)=\int_{-\infty}^{\infty}\td u\,e^{i E u}\tilde{F}(E)$, we have that
\begin{multline}
\label{dupu}
	\int_{-\infty}^{\infty}\td u\,\partial_uF(u)=2\pi i\lim_{E\to0}\left[E\tilde{F}(E)\right] \quad\implies\\
	\int_{-\infty}^{\infty}\td u\,\partial_u\bar\sigma^0=-\frac{1}{8\pi}\lim_{E_q\to0}\left[E_q\left(\mathfrak{a}_+(-E_q\hat x)-\mathfrak{a}^\dagger_-(E_q\hat x)\right)\right]\,.
\end{multline}
Regarding the linear piece, its action on asymptotic states is defined through canonical boundary commutation relations for the matter field~\cite{Frolov:1977bp,Lysov:2014csa}. More precisely, we have a boundary massless complex scalar $\phi(u,z,\bz)$ (\textit{cf.} equation~\eqref{Phir}), satisfying
\be
[\bar{\phi}(u,z,\bz),\phi(u',w,\bw)]=i\gawi\,\Theta(u'-u)\delta(z-w)\,.
\ee
From here, the following two commutation relations follow for the momentum-space field $\tilde{\phi}(E,w,\bw)=\int\td u\,e^{iuE}\phi(u,w,\bw)$:
\begin{equation}
\label{commus}
  \left[\partial_u\bar\phi,\tilde\phi\right]=-i\gai\delta(z-w)e^{iuE}\,,\qquad
  \left[\partial_z\bar\phi,\tilde\phi\right]=-\partial_z\left(\gai\delta(z-w)\right)\frac{e^{iuE}}{E}\,.
\end{equation}
Although $T^0_{uu}$ contains arbitrary integration constants (\textit{cf.} equation~\eqref{modify}), we can see in~\eqref{Qleading} that they drop out from the charge. Therefore, we have now spelled out all the ingredients needed to compute the action of the charges~\eqref{Q0NL} and~\eqref{Q0L} on our asymptotic states. The only thing left to specify is our choice for the supertranslation $f(z,\bz)$. As for the soft photon theorem, the preferred choice is \cite{He:2014laa}
\begin{equation}
\label{choosef}
  f(z,\bz)=\frac{1}{w-z}\,,
\end{equation}
where $w$ is identified with the direction on the sky of a soft graviton with negative helicity (the supertranslation for a positive-helicity one would be just the complex conjugate).
Combining all the ingredients we have just explained, we can immediately obtain:
\begin{gather}
\label{Q0NL2}
  \langle\textrm{out}|Q^{(0)}_{\textrm{NL}}|\textrm{in}\rangle=
  -\frac{1}{4\gaw}\,\p_{\bw}\left[\gaw\lim_{E_q\to0}
  \langle\textrm{out}|E_q\,\mathfrak{a}_+(-q)|\textrm{in}\rangle\right] \,,\\
\label{Q0L2}
	\langle\textrm{out}|Q^{(0)}_{\textrm{L}}|\textrm{in}\rangle=-\frac{1}{4}\frac{E_k}{w-w_k}\lim_{E_q\to0}
	\langle\textrm{out}|\textrm{in}\rangle \,.
\end{gather}
From the Ward identity~\eqref{Ward}, the form of the soft theorem that we get is
\begin{equation}
\label{D1soft}
  \p_{\bw}\left[\gaw\lim_{E_q\to0}\!\partial_{E_q}
  \langle\textrm{out}|E_q\,\mathfrak{a}_+(-q)|\textrm{in}\rangle\right]=
  -\gaw\sum_{k=1}^n\frac{E_k}{w-w_k}\lim_{E_q\to0}	\langle\textrm{out}|\textrm{in}\rangle\,.
\end{equation}
Although this looks superficially different from~\eqref{Dsoft-}, the two expressions are exactly the same. One just needs to manipulate $\frac{1+w\bw_k}{1+w_k\bw_k}=1+\frac{\bw_k(w-w_k)}{1+w_k\bw_k}$, and take into account from~\eqref{pk} that some components of the momentum conservation condition can be combined as
\begin{equation}
  \sum_{k=1}^n \left(p_{(k)1}-ip_{(k)2}\right)=\sum_{k=1}^n\frac{E_k \bw_k}{1+w_k\bw_k}=0 \,.
\end{equation}
Up to here, we have just simply re-derived the results already presented in~\cite{He:2014laa}. The good thing is that following the steps above for the sub-leading charges~\eqref{Qsub} and~\eqref{Qsubsub}, by essentially the same price we are going to obtain the sub- and sub-sub-leading soft graviton theorems.

\subsection{Sub-leading order}

Let us mimic the steps taken in the previous subsection. We start by stating the sub-leading soft graviton theorem:
\begin{equation}
\label{subsofth}
	\lim_{E_q\to0}\partial_{E_q}E_qM^{\pm}_{n+1}(p_1,\ldots,p_n,q)=
	\sum_{k=1}^n \frac{\epsilon^{\pm}\cdot p_k\,\epsilon^{\pm}_{\mu}q_{\nu}\,J_k^{\mu\nu}}{p_k\cdot q}\,M_{n}(p_1,\ldots,p_n)\,.
\end{equation}
where we have again used $\epsilon_{\mu\nu}^{\pm}(q)=\epsilon_{\mu}^{\pm}(q)\epsilon_{\nu}^{\pm}(q)$, and we should keep in mind that the right-hand side does not depend on $E_q$, although $q$ appears in there (the sub-leading soft factor just keeps information of the direction in which the graviton is taken soft). The angular momentum operator $J^{\mu\nu}$, acting on scalar particles, reads as:\footnote{
For massless particles in four dimensions, it is more convenient to write the angular momentum in helicity-spinor form. In this language, the corresponding soft factors can be found for instance in~\cite{Cachazo:2014fwa}.}
\begin{equation}
\label{Jmn}
J^{\mu\nu}_k=p_k^{\ \mu}\frac{\partial}{\partial p_{k\,\nu}}-p_k^{\ \nu}\frac{\partial}{\partial p_{k\,\mu}}\,.
\end{equation}
Now we translate~\eqref{subsofth} to null coordinates using~\eqref{qinw}-\eqref{einw}:
\begin{equation}
\label{subsoft+}
  \lim_{E_q\to0}\!\partial_{E_q}\!\langle\textrm{out}|E_q\,\mathfrak{a}_-(-q)|\textrm{in}\rangle=
  \sum_{k=1}^n \left(\frac{1+w\,\bw_k}{1+w_k\bw_k}\frac{w-w_k}{\bw-\bw_k}E_k\partial_{E_k}+
  \frac{(w-w_k)^2}{\bw-\bw_k}\partial_{w_k}\right)\langle\textrm{out}|\textrm{in}\rangle\,.
\end{equation}
Again with the benefit of hindsight, we have specified to a positive-helicity soft graviton, as this will be the natural choice associated to~\eqref{choosef}. The flip of helicity is not surprising as it was already encountered in the soft photon case~\cite{Conde:2016csj}. In there we also observed that the sub-leading order was naturally recovered from the sub-leading charge with two extra covariant derivatives respect to the leading order. It is then natural to act with three covariant derivatives on~\eqref{subsoft+}. In terms of partial derivatives, that is
\begin{multline}
\label{D3subsoft}
  \partial_w\,\gawi\partial_w\,\gawi\partial_w\left[\gaw\lim_{E_q\to0}\partial_{E_q}\langle\textrm{out}|E_q\,\mathfrak{a}_-(-q)|\textrm{in}\rangle\right]=\\
  2\pi\,\sum_{k=1}^n \left(-\gawik\left(\partial_{w_k}\delta\left(w-w_k\right)\right)E_k\partial_{E_k}+2\gawk\,\delta(w-w_k)\partial_{w_k}\right)\langle\textrm{out}|\textrm{in}\rangle\,,
\end{multline}
where we are denoting $\gawk\equiv\gamma_{w_k\bw_k}$ to avoid notation cluttering. To derive this formula above we needed to transform some $w$-derivatives into $w_k$-derivatives. This must be done with care because of the delta functions appearing.\footnote{
One should use that $f(x)\partial_x\delta(x-y)=-f(y)\partial_y\delta(x-y)-f'(y)\delta(x-y)$. More generally, if $O_z$ is an operator made of $n$ complex derivatives (say $\p_z$, $\p_{\bz}$), then $f(z)O_z[\delta(z-w)]=(-1)^nO_w[f(w)\delta(z-w)]$.}
Equation~\eqref{D3subsoft} is what we expect to obtain when we plug the sub-leading charge~\eqref{Qsub} in~\eqref{Ward}. To see that in detail, let us extract the relevant linear and non-linear pieces of~\eqref{Qsub}:
\begin{align}
\label{Q1NL}
 - Q^{(1)}_{\textrm{NL}}&=\frac13\int\td u\td z\td\bz\,\left(\p_z\p_{\bz}f\right)u\,\p_u\gai\p_z\gai\p_z\ga\sigma^0\,,\\
\label{Q1L}
  -Q^{(1)}_{\textrm{L}}&=\frac16\int\td u\td z\td\bz\left[\frac{u}{2}\left(\partial_z\partial_{\bz}f\right)T^0_{uu}+\left(\partial_{\bz}f\right)T^0_{uz}\right]\,.
\end{align}
To work out the action of the non-linear piece, this time it is convenient to use the following Fourier relation:
\begin{multline}
\label{udupu}
	\int_{-\infty}^{\infty}\td u\,u\,\partial_uF(u)=-2\pi \lim_{E\to0}\left[\partial_{E}\left(E\tilde{F}(E)\right)\right]\quad\implies\\
	\int_{-\infty}^{\infty}\td u\,u\,\partial_u\sigma^0=\frac{i}{16\pi}\lim_{E_q\to0}\left[\partial_{E_q}\,E_q\left(\mathfrak{a}_-(-E_q\hat x)-\mathfrak{a}^\dagger_+(E_q\hat x)\right)\right]\,.
\end{multline}
Then, for the choice~\eqref{choosef}, integrating once by parts we clearly obtain
\begin{equation}
\label{Q1NLf}
  \langle\textrm{out}|Q^{(1)}_{\textrm{NL}}|\textrm{in}\rangle=
  \frac{i}{12}\p_{w}\gawi\p_w\gawi\p_w\left[\gaw\lim_{E_q\to0}\!\partial_{E_q}
  \langle\textrm{out}|E_q\,\mathfrak{a}_-(-q)|\textrm{in}\rangle\right] \,.
\end{equation}
For the linear piece of the sub-leading piece, taking into account that again the integration constants present in the modified stress-energy tensor~\eqref{modify} disappear in the charge (as we can see in~\eqref{Qsub}), and using the relations~\eqref{commus} we immediately get
\begin{align}
\frac{1}{12}\int\td u\td z\td\bz\,u\left(\partial_z\partial_{\bz}f\right)\!\left[T^0_{uu},\tilde\phi_k\right]
&=-\frac{i\,\pi}{6}\gawik\left(\p_{w_k}\delta(w-w_k)\right)\left(1+E_k\partial_{E_k}\right)\tilde\phi_k\,,\\
\label{Q1Lf}
\int\!\td u\td z\td\bz\left(\partial_{\bz}f\right)\!\left[T^0_{uz},\tilde\phi_k\right]
&=\frac{i\,\pi}{6}\left(\gawik\left(\p_{w_k}\delta(w-w_k)\right)+2\gawik\delta(w-w_k)\partial_{w_k}\right)\tilde\phi_k\,,
\end{align}
where $\tilde\phi_k=\tilde\phi(E_k,w_k,\bw_k)$. It is straightforward to see that we nicely obtain~\eqref{D3subsoft} when we combine~\eqref{Q1NLf}-\eqref{Q1Lf} following~\eqref{Ward}.

\subsection{Sub-sub-leading order}

After having recovered the sub-leading soft graviton theorem from the charge~\eqref{Qsub}, it is not too surprising that we can also recover the sub-sub-leading one from the charge~\eqref{Qsubsub}. Still, it is very pleasing to see it explicitly as all the factors have to conspire properly. Let us show it here. At tree level in Einstein-Hilbert gravity we have
\begin{equation}
\label{subsubsofth}
	\lim_{E_q\to0}\partial^2_{E_q}E_qM^{\pm}_{n+1}(p_1,\ldots,p_n,q)=
	\sum_{k=1}^n \frac{\epsilon^{\pm}_{\mu}\epsilon^{\pm}_{\nu}q_{\rho}q_{\sigma}\,J_k^{\rho\mu}\,J_k^{\sigma\nu}}{E_q\,p_k\cdot q}M_{n}(p_1,\ldots,p_n)\,,
\end{equation}
where again we have used $\epsilon_{\mu\nu}^{\pm}(q)=\epsilon_{\mu}^{\pm}(q)\epsilon_{\nu}^{\pm}(q)$ and have deceivingly included $E_q$ on the right-hand side, which does actually not depend on it. The translation of~\eqref{subsubsofth} for a positive-helicity soft graviton to position-space coordinates, using~\eqref{qinw}-\eqref{einw}, yields
\begin{equation}
\label{subsubsoft+}
\begin{aligned}
  \lim_{E_q\to0}\partial^2_{E_q}\langle\textrm{out}|E_q\,\mathfrak{a}_-(-q)|\textrm{in}\rangle=&
  -\sum_{k=1}^n\frac{\bw-\bw_k}{w-w_k}\frac{1+w_k\bw_k}{1+w \bw}\bigg[\frac{(1+w\,\bw_k)^2}{(1+w_k\bw_k)^2}E_k\partial^2_{E_k}\\
  &\hspace{-25ex}+2(w-w_k)\left(\frac{1+w\,\bw_k}{1+w_k\bw_k}\,\partial^2_{E_k,w_k}-E_k^{-1}\partial_{w_k}\right)+(w-w_k)^2E_k^{-1}\partial^2_{w_k}\bigg]
  \langle\textrm{out}|\textrm{in}\rangle\,.
\end{aligned}
\end{equation}
Let us now judiciously take five derivatives on this expression.
\begin{multline}
\label{D5subsubsoft}
  \gawi\p_{\bw}\gaw\p_w\gawi\p_w\gawi\p_w\gawi\p_w\left[\gamma_w\langle\textrm{out}|\lim_{E_q\to0}\partial^2_{E_q}E_q\,\mathfrak{a}_-(-q)|\textrm{in}\rangle\right]=2\pi\,\times\\
 \times\sum_{k=1}^n\bigg[\left(\gawik\p_{w_k}\gawik\p_{w_k}\gaw\p_{\bw_k}\gawik\delta(w-w_k)\right)E_k\partial^2_{E_k}
 -4\left(\gawk^{-2}\p_{w_k}\gawk\p_{\bw_k}\gawik\delta(w-w_k)\right)\p_{E_k}\p_{w_k}\\
+\frac{6}{E_k}\left(\gawik\p_{\bw_k}\gawik\delta(w-w_k)\right)\p^2_{w_k}+\frac{4}{E_k}\left(\gawk^{-\frac12}\p_{w_k}\gawk^{-\frac12}\p_{\bw_k}\gawik\delta(w-w_k)\right)\partial_{w_k}\bigg]\langle\textrm{out}|\textrm{in}\rangle\,,
\end{multline}
In order to recover this expression from the sub-sub-leading charge~\eqref{Qsubsub}, we first rewrite its negative-helicity part in terms of partial derivatives. The result is a bit ugly, but practical:
\begin{align}
\label{Q2NL}
  Q^{(2)}_{\textrm{NL}}&=\frac{1}{24}\int\td u\td z\td\bz\,u^2\p_u\,\left(\pz\gai\pz\ga\pb\gai\pb f\right)\left(\gai\pz\gai\pz\ga\sigma^0\right)\,,\\
\nn
  Q^{(2)}_{\textrm{L}}&=\frac{1}{48}\int\td u\td z\td\bz\bigg[
  \frac{u^2}{2}\left(\pz\gai\pz\ga\pb\gai\pb f\right)T^0_{uu}\\
\label{Q2L}
  &\hspace{16ex}+2u\,\gai\left(\pz\ga\pb\gai\pz f\right)T^0_{uz}+
  3\left(\pb\gai\pb f\right)T^0_{zz}\bigg]\,.
\end{align}
The components of the boundary stress-energy tensor can be read from~\eqref{modify}. Unlike for the first two orders, leading and sub-leading, the integration constants $A_{\mu}$ do not cancel out in the charge. In order to recover~\eqref{subsubsofth}, we have to set them to zero. We believe these integration constants might be related to loop corrections, as we briefly mention in the conclusions. Now the sub-sub-leading charge is the same as the charge derived in~\cite{Campiglia:2016jdj,Campiglia:2016efb}. This is well expected as they both recover the sub-sub-leading soft graviton theorem~\eqref{subsubsofth} correctly.

Now let us work out the action of these charges on $\textit{out}$ states, starting with the non-linear piece. For that, the relevant Fourier relation is now
\begin{multline}
\label{u2dupu}
	\int_{-\infty}^{\infty}\td u\,u^2\,\partial_uF(u)=-2\pi i \lim_{E\to0}\left[\partial^2_{E}\left(E\tilde{F}(E)\right)\right]\quad\implies\\
	\int_{-\infty}^{\infty}\td u\,u^2\partial_u\sigma^0=\frac{1}{8\pi}\lim_{E_q\to0}\left[\partial^2_{E_q}\,E_q\left(\mathfrak{a}_-(-E_q\hat x)-\mathfrak{a}^\dagger_+(E_q\hat x)\right)\right]\,.
\end{multline}
Simply inserting~\eqref{choosef} in~\eqref{Q2NL}, and integrating thrice by parts we obtain:
\begin{equation}
\label{Q2NLf}
  \langle\textrm{out}|Q^{(2)}_{\textrm{NL}}|\textrm{in}\rangle=
  \frac{1}{96}\,\gawi\p_{\bw}\gaw\p_w\gawi\p_w\gawi\p_w\gawi\p_w \left[\gaw\lim_{E_q\to0}\!\partial^2_{E_q}
  \langle\textrm{out}|E_q\,\mathfrak{a}_-(-q)|\textrm{in}\rangle\right] \,.
\end{equation}
Next we lay out the action of each of the three terms in the linear piece~\eqref{Q2L}. From the one containing $T^0_{uu}$ we get:
\begin{multline}
\label{Tuu.act}
\frac{1}{96}\int\td u\td z\td\bz\left(\pz\gai\pz\ga\pb\gai\pb f\right)u^2\left[T^0_{uu},\tilde\phi_k\right]\\
=-\frac{\pi}{48}\left(\gawik\p_{w_k}\gawik\p_{w_k}\gawk\p_{\bw_k}\gawik\delta(w-w_k)\right)\left(E_k\partial^2_{E_k}+2\partial_{E_k}\right)\tilde\phi_k\,,
\end{multline}
while the piece with $T^0_{uz}$ yields
\begin{multline}
\label{Tuz.act}
\frac{1}{24}\int\td u\td z\td\bz\left(\pz\ga\pb\gai\pb f\right)u\left[T^0_{uz},\tilde\phi_k\right]
=\frac{\pi}{24}\bigg[2\left(\gawk^{-2}\p_{w_k}\gawk\p_{\bw_k}\gawik\delta(w-w_k)\right)\p_{E_k}\p_{w_k}\\
+\!\left(\gawik\p_{w_k}\gawik\p_{w_k}\gawk\p_{\bw_k}\gawik\delta(w-w_k)\right)\!\p_{E_k}
\!+\!E_k^{-1}\!\left(\gawk^{-2}\p_{w_k}\gawk\p_{\bw_k}\gawik\delta(w-w_k)\right)\!\p_{w_k}\bigg]\tilde\phi_k\,,
\end{multline}
and finally
\begin{multline}
\label{Tzz.act}
\frac{1}{16}\int\td u\td z\td\bz\left(\pb\gai\pb f\right)\left[T^0_{zz},\tilde\phi_k\right]\\
=-\frac{\pi}{8E_k}\bigg[\left(\gawik\p_{\bw_k}\gawik\delta(w-w_k)\right)\p^2_{w_k}
+\left(\gawik\p_{w_k}\p_{\bw_k}\gawik\delta(w-w_k)\right)\p_{w_k}\bigg]\tilde\phi_k\,.
\end{multline}
If we now combine in the Ward identity~\eqref{Ward} equations~\eqref{Q2NLf} with~\eqref{Tuu.act}-\eqref{Tzz.act}, using $6\gawik\p_{w_k}-2\gawk^{-2}\p_{w_k}\gawk=4\gawk^{-\frac12}\p_{w_k}\gawk^{-\frac12}$, we get exactly~\eqref{D5subsubsoft}!

\section{Discussion}

It is well established that the Ward identities associated to BMS supertranslations can be recast as Weinberg's leading soft graviton theorem~\cite{He:2014laa}. In this paper we have shown detailedly how BMS supertranslations are enough to reproduce the sub- and sub-sub-leading pieces of the soft graviton theorem recently discovered at tree level. In order to do that, we have argued that sub-leading orders (in the energy of the soft particle) in a soft theorem are in correspondence with sub-leading orders (in the inverse radial coordinate $\frac{1}{r}$) of asymptotic charges. This is an idea that we have already explored for the soft photon theorem in~\cite{Conde:2016csj}, where we also successfully obtained the sub-leading order of the theorem from the large gauge transformations that are behind the leading order. We believe that the present results on the soft graviton theorem provide a critical check of the consistency of the picture we propose.

Let us comment on some aspects that may cross the reader's mind. Maybe the most glaring one is that the $\frac{1}{r}$ expansion of the charge \eqref{charge'} goes much further than the sub-sub-leading order, so that from the correspondence we propose one would naively expect to discover soft graviton theorems beyond sub-sub-leading order. However, a careful look at the details presented in Section~\ref{ssec:asympQ} shows that this is not the case.
In order to recover the corresponding soft factors, the charges \eqref{Qleading}-\eqref{Qsubsub} are re-organized with several integrations by parts. Those integrations can only be performed in the absence of boundary contributions, which depends on the fall-off conditions at $\skyp_{\pm}$. The boundary condition we impose in this current work does not permit the desired arrangement of the charges beyond $\cO {r^{-2}}$. Hence, as a symmetry at null infinity, the supertranslation is insufficient to yield universal properties of soft emission beyond sub-sub-leading order. From a different point of view, we may see this as the counterpart of the fact that gauge invariance only fixes three orders of soft emission in terms of the non-radiative amplitude (see for instance~\cite{Bern:2014vva}).

One may wonder if the sub-leading orders of the charge \eqref{charge'} can match the sub-leading pieces of the standard expression for asymptotic conserved charge in \cite{Wald:1999wa,Barnich:2001jy} which does lead to a sub-leading analysis in lower dimensional gravity \cite{Compere:2014cna} (see also \cite{Compere:2015knw}). A direct computation shows that it is not the case. Alternatively, one can consider the soft theorem as a prescription to define the charge uniquely. Nonetheless, it would be very meaningful to explore the physical nature of the charge \eqref{charge'} elsewhere.

A more subtle point has to do with the observation that, at sub-sub-leading order~\eqref{Qsubsub}, our charge $Q^{(2)}$ is sensitive to the integration constants that naturally appear when shifting the stress-energy tensor, as required by the consistency of our method. These integration constants could \textit{a priori} also appear in $Q^{(0)}$ and $Q^{(1)}$, but stunningly they do not. A similar phenomenon was happening in the case of the soft photon theorem, with integration constants involved in the shifted current showing up only at sub-leading order~\cite{Conde:2016csj}. Since these are precisely the orders at which the loop corrections enter the game, we suspect there could be a connection between the two phenomena. A possible way of investigating this further would be to consider non-minimal couplings of matter to gravitons (recently considered in~\cite{Elvang:2016qvq}) from the phase-space perspective of the current work, and see if the aforementioned integration constants play a role there. We leave that for future investigation.

A puzzling issue to which we do not have currently an answer is what is the fate of the Virasoro algebra that comes with superrotations, and which would be presumably important for recent attempts at flat-space holography~\cite{Kapec:2016jld,Cheung:2016iub}. It may be that the idea of associating the Virasoro algebra of the would-be dual conformal field theory to that of superrotations is simply too naive, and the way this algebra should be encoded is just more involved.

Finally, another direction of future research could be the following: Given that the picture we are proposing for sub-leading orders of soft theorems seems to be consistent, an obvious case for which such analysis is missing is that of Yang-Mills theory~\cite{Casali:2014xpa,White:2014qia}, where the connection with asymptotic symmetries is actually richer.
The leading soft theorem for single soft gluon emission can be linked to certain large gauge transformations~\cite{He:2015zea}, similarly to its Abelian analogue, Maxwell theory. Naively it is straightforward to extend to Yang-Mills theory our proposal of connecting the sub-leading soft theorem to the sub-leading structure of null infinity, given the fact that such connection was well established in the Abelian case~\cite{Conde:2016csj}. However, a very novel property of the soft theorem in the non-Abelian theory is that the soft factor does not obey a simple factorization in multiple soft gluon emission even at the leading piece \cite{Berends:1988zn}, namely the soft factor does see the order of the gluons taken to be soft. Such phenomenon is conjectured to be reflecting the structure of the symmetry group controlling the soft theorem \cite{ArkaniHamed:2008gz}.
It would be very interesting to reveal those issues from the asymptotic symmetry point of view concretely elsewhere \cite{2come}.

\section*{Acknowledgments}
\label{sec:acknowledgements}


It is a pleasure to thank Glenn Barnich, Euihun Joung and Victor Lekeu for discussions which helped us clarify some aspects of this work. We would also like to thank Hongbao Zhang and the Department of Physics of Beijing Normal University for hospitality during a productive two weeks spent in Beijing, where part of this work was completed. EC also acknowledges CFGS at Seoul for hospitality during many autumn afternoons. The work of EC and PM are respectively supported by the National Research Foundation of Korea through the grant NRF-2014R1A6A3A04056670, and in part by NSFC Grant No. 11575202.

\appendix

\section{Modification of the stress-energy tensor}\label{solutionspace}

The stress-energy tensor $\tT^{\mu\nu}$ can be modified by a four-tensor $K^{\mu\sigma\nu\rho}$ with the following symmetries
\be
K^{\mu\sigma\nu\rho}=-K^{\sigma\mu\nu\rho}\,,\;\;\;\;K^{\mu\sigma\nu\rho}=-K^{\mu\sigma\rho\nu}\,\;\;\;\;K^{\mu\sigma\nu\rho}=K^{\nu\rho\mu\sigma}\,,
\ee
which guarantee that the expression $\nabla_\mu\nabla_\nu K^{\mu\sigma\nu\rho}$ is conserved automatically and symmetric in the indices $\sigma$ and $\rho$. This last property follows from the fact that the connection is flat, namely $[\cd_\mu,\cd_\nu]=0$. Hence, we can modify the stress tensor by
\be
T^{\mu\nu}={\tT}^{\mu\nu}-\cd_\sigma\cd_\rho K^{\mu\sigma\nu\rho}\,.\label{shift}
\ee
A simple, but not necessarily the only choice,\footnote{
Although the choice of shift is not unique, the leading order of the any shifted stress-energy momentum satisfying ${T}_{r\alpha}=0$ has always the same form.
}
of this $K$ tensor to set ${T}^{u\alpha}=0$, or equivalently ${T}_{r\alpha}=0$, would be
\be\label{K}\begin{split}
&K^{urrz}=-\frac{1}{r}\int^{\infty}_r\;\frac{\td r'}{r'^3}\int^{\infty}_{r'}\;\td r''\tT^{uz}r''^4 - \frac{A_z(u,z,\bz)}{2r^3\ga}\,,\\
&K^{urr\bz}=-\frac{1}{r}\int^{\infty}_r\;\frac{\td r'}{r'^3}\int^{\infty}_{r'}\;\td r''\tT^{u\bz}r''^4 - \frac{A_{\bz}(u,z,\bz)}{2r^3\ga}\,,\\
&K^{urur}=\frac{1}{r^2}\int^{\infty}_r\;\td r'\int^{\infty}_{r'}\;\td r''\tT^{uu}r''^2 - \frac{A_r(u,z,\bz)}{r^2}\,,\\
&K^{rzr\bz}=\frac{1}{2r^3\ga^2}\int^{\infty}_r\;\td r'\bigg[\ga r'^2 \tT^{ur}- 3 r \pb (\ga K^{urr\bz}) - 3r \p_z (\ga K^{urrz})\\
&\hspace{1.5cm}- r^2\pb\p_r(\ga K^{urr\bz}) - r^2 \p_z\p_r(\ga K^{urrz}) + \ga\p_r(r^2K^{urrz})\bigg] - \frac{A_u(u,z,\bz)}{r^3\ga}\,,
\end{split}\ee
where $A_z(u,z,\bz),A_{\bz}(u,z,\bz),A_r(u,z,\bz),A_u(u,z,\bz)$ are arbitrary integration constants. The remaining components of $K$ are zero.

We are interested in coupling complex scalar matter to the gravitational theory. The conservation law of the stress-energy tensor $\cd_\mu \tT^{\mu\nu}=0$ is guaranteed when the Klein-Gordon equation is satisfied. The solution of those equations was well demonstrated in \cite{Bondi:1962px} as the first example of the Bondi-Sachs framework. Once the initial data of the scalar fields are given in series expansion of $\frac1r$,
\be\label{Phir}\begin{split}
&\Phi(u,r,z,\bz)=\frac{\phi(u,z,\bz)}{r}+\sum\limits_{m=1}^\infty\frac{\Phi^{(m)}(u,z,\bz)}{r^{m+1}}\,,\\
&\xbar\Phi(u,r,z,\bz)=\frac{\xbar\phi(u,z,\bz)}{r}+\sum\limits_{m=1}^\infty\frac{\xbar\Phi^{(m)}(u,z,\bz)}{r^{m+1}}\,.
\end{split}\ee
the Klein-Gordon equation will yield the following results: \textit{$\phi$ and $\xbar\phi$ are news function reflecting the radiative part of the fields and the time evolution of all $\Phi^{(m)}$ terms will be controlled.} The stress-energy tensor of a complex scalar is
\be\nn\begin{split}
&\tT_{uu}=\p_u\Phi\,\p_u\xbar\Phi\,,\\
&\tT_{uz}=\half\p_u\Phi\,\p_z\xbar\Phi+\half\p_u\xbar\Phi\,\p_z\Phi\,,\\
&\tT_{zz}^0=\p_z\Phi\p_z\xbar\Phi\,.\end{split}
\ee
By implementing \eqref{shift} with the precise choice of $K$ in \eqref{K}, the leading piece of the modified stress-energy tensor is obtained as:
\be\begin{split}
&T_{uu}^0=\p_u\phi\,\p_u\xbar\phi + 2\p_u A_u + \p_u^2 A_r\,,\\
&T_{uz}^0=\p_u A_{\bz} - \p_z A_u-\half\phi\,\p_u\p_z\xbar\phi-\half\xbar\phi\,\p_u\p_z\phi\,,\\
&T_{zz}^0=\p_z\phi\,\p_z\xbar\phi\,.\end{split}\label{modify}
\ee

\section{Spin-weighted derivative operators}
\label{eth}

The definitions of $\eth$ and $\xbar\eth$ on a field $\eta$ with spin weight $s$ are
\be
\eth \eta=\ga^{\frac{s-1}{2}}\pzb (\eta\ga^{-\frac{s}{2}})\,,\;\;\;\; \xbar\eth \eta=\ga^{\frac{-s-1}{2}}\pz (\eta\ga^{\frac{s}{2}})\,,
\ee
and the spin weight of the relevant fields can be found in Table \ref{t1}.
\begin{table}[h]
\caption{Spin weights}\label{t1}
\begin{center}
\begin{tabular}{c|c|c|c|c|c|c|c|c|c|c} & $\sigma^0$  & $\dot\sigma^0$ &
   $\Psi^0_3$ &$\Psi^0_2$ & $\Psi^0_1$ &$\Psi_0^0$ & $\Psi_0^{(m)}$  & $\cY$ \\
\hline
s &  $2$ &  $2$&$-1$  & $0$ & $1$ & $2$  & $2$ & $-1$\\
\end{tabular}
\end{center} \end{table}
Notice that $\eth$ ($\xbar\eth$) increases (decreases) the spin weight. These two derivative operators do not commute in general. The commutation relation is
\be
[\xbar\eth,\eth]\eta=s\eta\,,
\ee
where $s$ is the spin weight of the field $\eta$. The actions of $\eth$ and $\xbar\eth$ on the spherical harmonics $Y_{l,m}$ $(l=0,1,2,\ldots\,;\,m=-l,\ldots,l)$ are very useful for calculations on the sphere. Spin $s$ spherical harmonics are defined as
\begin{equation}
	\Preind_{s}{Y}_{l,m}=\begin{cases}
	\sqrt{\dfrac{(l-s)!}{2(l+s)!}}\,\eth^s Y_{l,m}\qquad\qquad(0\leq s\leq l)\\
	(-1)^s\sqrt{\dfrac{(l+s)!}{2(l-s)!}}\,\xbar\eth^{-s} Y_{l,m}\quad(-l\leq s\leq 0)
\end{cases} \,.
\end{equation}
From there, one can find that
\begin{equation}
\label{zero}
\begin{gathered}
	\int\td z \td\bz\,\gamma_{z\bz}\, \Preind_{s}{Y}_{l,m}\,\eth^{l-s+1}\eta =0 \,,\qquad
	\int \td z \td\bz\,\gamma_{z\bz}\,\Preind_{s}{\bar{Y}}_{l,m}  \,\xbar\eth^{l-s+1}\zeta=0 \,,\\
	\xbar\eth\eth\Preind_{s}{Y}_{l,m}=-\half(l-s)(l+s+1)\Preind_{s}{Y}_{l,m} \,, \qquad
	\int\td z \td\bz\,\gamma_{z\bz}\, A\eth B=-\int\td z \td\bz\,\gamma_{z\bz}\, B\eth A \,.
\end{gathered}
\end{equation}
where $\eta$ and $\zeta$ have spin weight $-l-1$ and $l+1$ respectively, and the expression $A\eth B$ should have spin weight zero.

\section{Newman-Penrose charges}
\label{NP}

We briefly review the main results of the linearized gravitational theory part in~\cite{Newman:1968uj} in this appendix. The Newman-Penrose (first-order) formalism~\cite{Newman:1961qr} was used in that work. The Newman-Penrose equations yield the following relations:
\be\label{NPresults}
\begin{split}
\p_u\Psi_2^0&=-\eth\Psi_3^0\,,\\
\p_u\Psi_1^0&=-\eth\Psi_2^0\,,\\
\p_u\Psi_0^0&=-\eth\Psi_1^0\,,\\
(n+1)\p_u\Psi_0^{n+1}&=-\big(\xbar\eth\eth+(n+5)n\big)\Psi_0^n\;\;\;\;(n>0)\,.
\end{split}\ee
The first equation of \eqref{NPresults} leads to the conservation of mass
\be
\p_u\int\td z \td\bz\,\gamma_{z\bz}\;\Preind_{0}{Y}_{0,0}\,\Psi_2^0=-\int\td z \td\bz\,\gamma_{z\bz}\;\Preind_{0}{Y}_{0,0}\,\eth\Psi_3^0=0\,,\label{masslaw}
\ee
where the relations in \eqref{zero} have been used. The novel discovery of \cite{Newman:1968uj} was the infinite amount of new conservation laws from the last equation of \eqref{NPresults}, namely
\be\begin{split}
(n+1)\p_u\int\td z \td\bz\,\gamma_{z\bz}\;\Preind_{2}{Y}_{n-k+2,m}\,\Psi_0^{n+1}&=-\int\td z \td\bz\,\gamma_{z\bz}\;\Preind_{2}{Y}_{n-k+2,m}\,\big(\xbar\eth\eth+(n+5)n\big)\Psi_0^n\\
&=-\int\td z \td\bz\,\gamma_{z\bz}\;\Preind_{2}{Y}_{n-k+2,m}\,k(2n-k+5)\Psi_0^n\,,
\end{split}\ee
whence
\be
\p_u\int\td z \td\bz\,\gamma_{z\bz}\;\Preind_{2}{Y}_{n+2,m}\,\Psi_0^{n+1}=0\,.\label{newlaws}
\ee
Meanwhile the second and third equations of \eqref{NPresults} can only induce identically vanishing quantities
\begin{align}
\p_u\int\td z \td\bz\,\gamma_{z\bz}\;\Preind_{0}{Y}_{0,0}\,\xbar\eth\Psi_1^0&=-\int\td z \td\bz\,\gamma_{z\bz}\;\Preind_{0}{Y}_{0,0}\,\xbar\eth\eth\Psi_2^0=0\,,\label{trivial1}\\
\p_u\int\td z \td\bz\,\gamma_{z\bz}\;\Preind_{0}{Y}_{0,0}\,\xbar\eth^2\Psi_0^0&=-\int\td z \td\bz\,\gamma_{z\bz}\;\Preind_{0}{Y}_{0,0}\,\xbar\eth^2\eth\Psi_1^0=0\,.\label{trivial2}
\end{align}

\bibliography{amplitudesrefs,asymptrefs,local}
\bibliographystyle{utphys}

\end{document}